\documentclass[twocolumn,showpacs,pra]{revtex4}%
\usepackage{amssymb}
\usepackage{graphicx,color,psfrag}
\usepackage{dcolumn}
\usepackage{bm}
\usepackage{amsmath}
\usepackage{amsfonts}
\begin{document}
\title{Scattering Resonances in a Degenerate Fermi Gas}
\author{K. J. Challis}
\author{N. Nygaard}
\author{K. M{\o}lmer}
\affiliation{Lundbeck Foundation Theoretical Center for Quantum System Research, \\
Department of Physics and Astronomy, University of Aarhus, DK-8000 $\AA$rhus C, Denmark}
\date{\today}

\begin{abstract}
We consider elastic single-particle scattering from a one-dimensional two-component trapped superfluid Fermi gas when the incoming projectile particle is identical to one of the confined species.  Our theoretical treatment is based on the Hartree-Fock ground state of the trapped gas and a configuration-interaction description of the excitations.  We determine the scattering phase shifts for the system and predict Fano-type scattering resonances that are a direct consequence of inter-atomic pairing.  We describe the main characteristics of the scattering resonances and make a comparison with the results of BCS mean-field  theory.
\end{abstract}

\pacs{03.75.Ss, 03.65.Nk}
\maketitle

Scattering experiments have led to many important discoveries in physics.  In the field of dilute atomic gases, an understanding of inter-atomic scattering, as well as scattering between light and atoms, is fundamental to our ability to prepare, control, and manipulate systems for study.  For example, a Feshbach resonance is a scattering resonance that makes it possible to control the strength and sign of inter-atomic interactions \cite{Inouye98}.  An understanding of this phenomenon has led to the formation of molecular condensates \cite{Jochim03, Regal03} and the investigation of the crossover between Fermi superfluidity and Bose-Einstein condensation \cite{Zwierlein05, Chin04}.

In this paper we consider elastic single-particle scattering from a one-dimensional two-component degenerate Fermi gas when the projectile particle is identical to one of the confined species.  Our theoretical treatment is a number-conserving approach based on the Hartree-Fock ground state of the trapped gas.  We construct the excitations in a configuration-interaction approach using one-particle continuum excitations and bound two-particle one-hole excitations.  The excitations are single-particle-like in the asymptotic limit and by extracting the scattering phase shifts we determine the scattering properties of the system.

We predict Fano-type scattering resonances that arise due to the interrelation of the one- and multi-particle branches of the excitation spectrum.  In particular, when the energy of a two-particle one-hole bound state lies within the one-particle continuum, coupling between the two branches significantly modifies the scattering properties of the system near the uncoupled bound state energy.  In this system, the two branches of the excitation spectrum are coupled by the inter-atomic pairing.

We also present a BCS mean-field description of the scattering.  In this approach, the scattering properties of the system are determined from the asymptotic behavior of the many-body quasiparticle excitations.  We find that, in the uncoupled system, bound hole excitations lie within the particle excitation continuum.  The mean-field pair potential couples the particle and hole branches of the excitation spectrum and we again observe Fano-type scattering resonances.  However, these resonances are quantitatively different from those predicted by the configuration-interaction method.  In this particular system, the BCS treatment leads to spurious results that are explicitly linked with the violation of particle number conservation in the theory. Effects of this type are particularly evident in small systems.

Our paper is organized as follows.  In Sec.\ \ref{sec:intro} we outline the system for study.  In Sec.\ \ref{sec:HF} we investigate elastic single-particle scattering from a trapped Fermi gas using the configuration-interaction method.  In Sec.\ \ref{sec:BCS} we present the BCS mean-field treatment of that same problem.  In Sec.\ \ref{sec:disc} we give a detailed discussion of the validity of the mean-field approach, and we conclude in Sec.\ \ref{sec:conclusion}.

\section{Two-component Fermi gas \label{sec:intro}}

We consider a trapped one-dimensional degenerate Fermi gas at zero temperature with two equally populated spin components interacting via an attractive contact potential.  The Hamiltonian is
\begin{equation}
\begin{split}
\hat{H} & = \sum_{\alpha} \int  \hat{\psi}_{\alpha}^{\dagger}(x) H_{\rm SP}(x) \hat{\psi}_{\alpha} (x) dx \\
& \quad + g \int \hat{\psi}_{\uparrow}^{\dagger}(x) \hat{\psi}_{\downarrow}^{\dagger}(x) \hat{\psi}_{\downarrow}(x) \hat{\psi}_{\uparrow} (x)dx,
\label{Hamiltonian}
\end{split}
\end{equation}
where the field operator $\hat{\psi}_{\alpha}(x)$ destroys a particle at position $x$, in the spin state $\alpha = \uparrow, \downarrow$, and $g<0$.  The single-particle Hamiltonian is
\begin{equation}
H_{\rm SP}(x) = -\frac{\hbar^2}{2M} \frac{d^2}{dx^2} + V_{\rm ext} (x),
\end{equation}
where $M$ is the atomic mass and $V_{\rm ext}(x)$ is a symmetric external trapping potential with a zero energy continuum threshold.  The exact form of the trapping potential is not crucial for our discussion.  However, for our numerical calculations we use the Gaussian form $V_{\rm ext}(x)= -V_0 \exp (-2x^2/w^2)$, with $V_0>0$, which can be realized for atoms in a waveguide by applying a single Gaussian laser beam.  The Gaussian width provides a convenient energy scale $E_w=\hbar^2/2Mw^2$.

\section{Configuration-interaction method \label{sec:HF}}

To describe elastic single-particle scattering we determine the excitation spectrum using the equation of motion method \cite{R&S}.  The ground state $|{\rm G}\rangle$ of the trapped gas has energy $E_{\rm G}$, i.e., $\hat{H}|{\rm G}\rangle=E_{\rm G}|{\rm G}\rangle.$  We introduce the operators $\hat{Q}_{\nu}^{\dagger}$ that create excitations $|\nu \rangle = \hat{Q}_{\nu}^{\dagger}|{\rm G}\rangle$ satisfying $\hat{H}|\nu \rangle = E_{\nu}|\nu \rangle$, and the excitation spectrum is given by
\begin{equation}
[\hat{H}, \hat{Q}_{\nu}^{\dagger}] |{\rm G}\rangle = \bar{E}_{\nu} \hat{Q}_{\nu}^{\dagger} |{\rm G}\rangle,
\label{eqn_motion}
\end{equation}
where $\bar{E}_{\nu}=E_{\nu}-E_{\rm G}$.  Projecting Eq.\ (\ref{eqn_motion}) onto the state $\hat{X}^{\dagger}|{\rm G}\rangle$, where $\hat{X}$ is an arbitrary operator, gives
\begin{equation}
\langle {\rm G}| \hat{X} [\hat{H}, \hat{Q}_{\nu}^{\dagger}] |{\rm G}\rangle = \bar{E}_{\nu}\langle {\rm G}| \hat{X}  \hat{Q}_{\nu}^{\dagger} |{\rm G}\rangle.
\label{evalue_eqn}
\end{equation}
Later, $\hat{X}$ is chosen so that Eq.\ (\ref{evalue_eqn}) generates a matrix eigenvalue equation for the excited states $|\nu\rangle$.  Equation (\ref{evalue_eqn}) is exact but is difficult to solve because we do not know the ground state $|{\rm G}\rangle$ or the excitation creation operators $\hat{Q}_{\nu}^{\dagger}$.

An approximate solution to Eq.\ (\ref{evalue_eqn}) can be found by neglecting pair correlations in the ground state, i.e., we calculate the matrix elements using the Hartree-Fock ground state $|{\rm HF}\rangle$ in place of $|{\rm G}\rangle$.  We briefly summarize the Hartree-Fock method here.  The Hartree-Fock Hamiltonian is
\begin{equation}
\hat{H}_{\rm HF} = \sum_{\alpha} \int  \hat{\psi}_{\alpha}^{\dagger}(x) \left[H_{\rm SP}(x) +W(x) \right] \hat{\psi}_{\alpha} (x) dx,
\label{Hamiltonian_HF}
\end{equation}
where $W(x) = g\langle{\rm HF}| \hat{\psi}_{\alpha}^{\dagger}(x) \hat{\psi}_{\alpha}(x)|{\rm HF}\rangle$ is spin independent because we have chosen equal spin populations.  The expansion $\hat{\psi}_{\alpha}(x) = \sum_n \phi_n(x) \hat{a}_{n\alpha}$ diagonalizes Hamiltonian (\ref{Hamiltonian_HF}) where the Hartree-Fock wavefunctions satisfy
\begin{equation}
\left[ H_{\rm SP}(x)+W(x) \right] \phi_n(x) = E_n \phi_n(x).
\label{HF_basis}
\end{equation}
The operators $\hat{a}_{n\alpha}^{\dagger}$ and $\hat{a}_{n\alpha}$ obey fermionic commutation relations and the modes are populated according to
\begin{equation}
\langle {\rm HF }| {\hat a}_{n\alpha}^{\dagger} {\hat a}_{n' \beta} |{\rm HF}\rangle = n_n  \delta_{nn'}\delta_{\alpha\beta},
\label{mode_popn}
\end{equation}
where, at zero temperature, $n_n=1$ for $E_n<E_{\rm F}$ and $n_n=0$ for $E_n>E_{\rm F}$.  We choose the Fermi energy $E_{\rm F}<0$ so that all of the particles are bound.  The Hartree-Fock ground state is constructed by adding one particle of each spin to the lowest available energy level until the Fermi energy is reached, i.e.,
\begin{equation}
|{\rm HF}\rangle = \left( \prod _{n \leq n_{\rm F}} \hat{a}^{\dagger}_{n\uparrow}\hat{a}^{\dagger}_{n\downarrow} \right)|0\rangle,
\label{HF_ground}
\end{equation}
where $n_{\rm F}$ denotes the highest occupied level.

\begin{figure}[t]
\includegraphics[width=8.5cm]{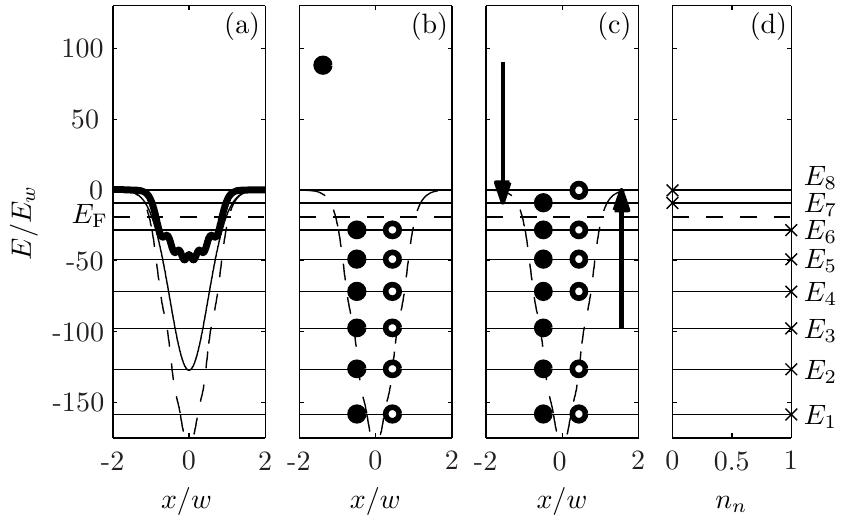}
\caption{(a) The Hartree-Fock ground state of a trapped Fermi gas.  The curves correspond to (thick) the Hartree potential $W(x)$, (thin) the Gaussian external potential $V_{\rm ext}(x)$, and (dashed) the combined potential $V_{\rm ext}(x)+W(x)$.  (b) A scattering one-particle excitation, and (c) a bound two-particle one-hole excitation,  where the energy level occupation of the ($\bullet$) spin up and ($\circ$) spin down particles is indicated schematically.  The arrows indicate the particle rearrangement between the configurations in panels (b) and (c).  (d) The ground state energy level occupation [see Eq.\ (\ref{mode_popn})] is indicated by the horizontal component of the markers $\times$.  In all panels, the horizontal lines indicate (dashed) the Fermi energy $E_{\rm F}$ and (solid) the bound Hartree-Fock energies $E_n<0$ [see Eq.\ (\ref{HF_basis})].  Parameters are $g = -9.55 w E_w$, $V_0 = 127.32 E_w$, and $E_{\rm F} = -19.10 E_w$. }
\label{fig:HF_ground}
\end{figure}

We determine the Hartree-Fock ground state of the system by numerically calculating the self-consistent Hartree potential $W(x)$.  We use an iterative method where the kinetic energy term in Eq.\ (\ref{HF_basis}) is evaluated according to a finite-difference formula.  Figure \ref{fig:HF_ground}(a) shows the Hartree potential for a particular set of parameters where there are six particles of each spin state, i.e., $\langle \hat{N}_{\alpha}\rangle=\int\langle \hat{\psi}_{\alpha}^{\dagger}(x) \hat{\psi}_{\alpha}(x) \rangle dx=6.00$. We consider a small number of particles so that the breakdown of BCS theory can be demonstrated clearly in Sec.\ \ref{sec:BCS}.  The small particle number means that there are oscillations in the Hartree potential in Fig.\ \ref{fig:HF_ground}(a) \cite{Bruun98}.  In the ground state the particles occupy the lowest available energy levels as shown in Fig.\ \ref{fig:HF_ground}(d) [see Eqs.\ (\ref{mode_popn}) and (\ref{HF_ground})].

We consider excitations $|\nu \rangle$ that are single-particle-like in the asymptotic limit but are phase shifted from plane-wave scattering states due to the presence of the external trapping potential and the trapped superfluid gas.  Formally, the operators $\hat{Q}_{\nu}^{\dagger}$ are constructed from continuum one-particle excitations and bound multi-particle excitations.  The multi-particle excitations in general consist of all possible configurations of the particles in the Hartree-Fock energy levels.  We include only the leading-order terms, i.e.,
\begin{equation}
\hat{Q}_\nu ^{\dagger}  = \sum_q C_q^{\nu} \hat{a}_{q \uparrow}^{\dagger} + \sum_{r,s,t} B_{rst}^{\nu} \hat{a}_{r \uparrow}^{\dagger} \hat{a}_{ s\downarrow}^{\dagger} \hat{a}_{t \downarrow},
\label{Q_excite}
\end{equation}
where $E_q>0$, and $E_r, E_s, E_t<0$ \cite{note_CImodes}.  The first term in Eq.\ (\ref{Q_excite}) describes continuum one-particle excitations [e.g., see Fig.\ \ref{fig:HF_ground}(b)].  The second term introduces bound two-particle one-hole excitations where the spin of the hole differs from the spin of the projectile particle [e.g., the bound excitation with $(r,s,t)=(7,8,3)$ is illustrated in Fig.\ \ref{fig:HF_ground}(c)].  We require $E_r, E_s > E_{\rm F}$ and $E_t<E_{\rm F}$ as these are the only non-zero contributions when acting on the Hartree-Fock ground state (\ref{HF_ground}).

The coefficients $C_q^{\nu}$ and $B_{rst}^{\nu}$ in Eq.\ (\ref{Q_excite}) are determined so that Eq.\ (\ref{evalue_eqn}) is valid, with $|{\rm G}\rangle \rightarrow |{\rm HF}\rangle$.  Taking $\hat{X}^{\dagger}= \hat{a}_{q\uparrow}^{\dagger}$ and $\hat{X}^{\dagger}= \hat{a}_{r \uparrow}^{\dagger} \hat{a}_{ s\downarrow}^{\dagger} \hat{a}_{t \downarrow}$ yields
\begin{equation}
 E_q C_q ^{\nu}  +\sum_{r,s,t}B_{rs t}^{\nu}V_{qtsr} = \bar{E}_{\nu} C_q^{\nu},
\label{eqn1_c}
\end{equation}
and
\begin{equation}
\begin{split}
 (E_r+E_s-E_t) B_{rst}^{\nu}+ \sum_q C_q^{\nu} V_{rstq}& \\
 + \sum_{r', s'} B_{r's't}^{\nu} V_{rss'r'}   - \sum_{r', t'} B_{r'st'}^{\nu}V_{rt'tr'} & = \bar{E}_{\nu} B_{rst}^{\nu},
 \end{split}
\label{eqn2_c}
\end{equation}
respectively, where
\begin{equation}
V_{n m m' n'} = g\int \phi_n^*(x) \phi_m ^* (x) \phi_{m'} (x) \phi_{n'} (x) dx.
\label{matrix_element}
\end{equation}
Equations (\ref{eqn1_c}) and (\ref{eqn2_c}) define a Hermitian eigenvalue problem for the coefficients $C_q^{\nu}$ and $B_{rst}^{\nu}$.  The one-particle continuum excitations $(C_q^{\nu})$ are coupled to the bound two-particle one-hole excitations $(B_{rst}^{\nu})$ by the full Hamiltonian (\ref{Hamiltonian}) giving rise to the off-diagonal interaction terms $V_{qtsr}.$   By construction, the three-particle wavefunctions associated with the coefficients $B_{rst}^{\nu}$ vanish asymptotically and, in the limit $x\rightarrow \pm \infty$, the excitations $|\nu\rangle$ are described by the single-particle wavefunctions
\begin{equation}
\Phi_{\nu}(x) = \sum_q C_q^{\nu} \phi_q (x).
\end{equation}

To investigate the scattering properties quantitatively, the Hartree-Fock ground state is computed on a grid of spatial extent $2L$.  We invoke the Neumann boundary conditions $[d \phi_n(x) / dx]_{x=\pm L} = 0$ so that the Hartree-Fock wavefunctions provide an orthonormal basis.  Solving the eigenvalue problem defined by Eqs.\ (\ref{eqn1_c}) and (\ref{eqn2_c}), we determine the single-particle wavefunctions $\Phi_{\nu}(x)$ for particular discrete values of the energy eigenvalue $\bar{E}_{\nu}$.  The scattering information at any energy of interest is then extracted using the R-matrix method \cite{Wigner47}, i.e., we match $\Phi_{\nu}(x)$ to the analytic asymptotic forms
\begin{eqnarray}
 \lim_{x \to \pm \infty} \Phi_{\nu}^{\rm{e}}(x)& \propto & \cos(kx \mp \delta_{\rm e} (k)) \nonumber \\
 \lim_{x\to \pm \infty} \Phi_{\nu}^{\rm{o}}(x)& \propto & \sin(kx \mp \delta_{\rm o} (k)),
\label{phi_asympt}
\end{eqnarray}
where $\bar{E}_{\nu}=\hbar^2k^2/2M$, and we use the Hartree-Fock basis to reconstruct the asymptotic solutions, and the scattering phase shifts $\delta_{\rm e}(k)$ and $\delta_{\rm o}(k)$, for any $k$.

\begin{figure}[t]
\includegraphics[width=8.5cm]{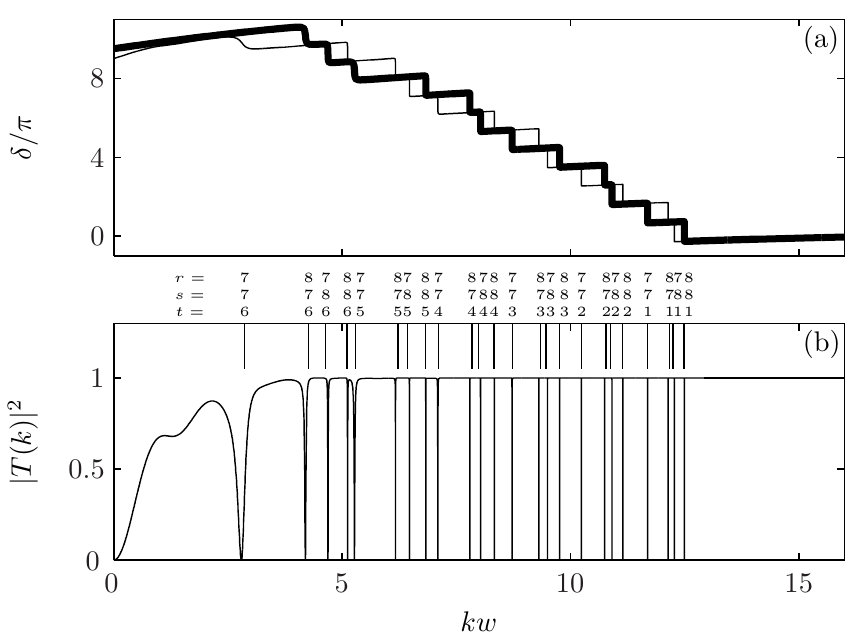}
\caption{(a) Scattering phase shifts and (b) transmission probability for a trapped Fermi gas, calculated using the configuration-interaction method.  The curves in (a) correspond to (thick) the even phase shift $\delta_{\rm e}(k)$ and (thin) the odd phase shift $\delta_{\rm o}(k)$.  The vertical lines in $(b)$ indicate the uncoupled resonance wavevectors $k_{rs}^{t}$ [see Eq.\ (\ref{Erst})].  Parameters are $g = -9.55 wE_w$, $V_0 = 127.32 E_w$, and $E_{\rm F} = -19.10 E_w$. }
\label{fig:CI_scat}
\end{figure}

The even and the odd phase shifts for the trapped Fermi gas considered in Fig.\ \ref{fig:HF_ground} are shown in Fig.\ \ref{fig:CI_scat}(a).  We observe a variation in the background phase shifts $\delta_{\rm e,o}^{\rm bg}(k)$  due to the effective external potential $V_{\rm ext}(x)+W(x)$ [see Eq.\ (\ref{HF_basis})].  We also observe twenty-four resonance features that are characterized by a jump of $\sim\pi$ in either the even or the odd phase shift.

Transmission and reflection coefficients for the system can be calculated from the phase shifts \cite{Poulsen03, Grupp06}.  For a projectile particle incident from $x=-\infty$ (or equivalently from $x=\infty$), the transmission probability (normalized to one) is given by $|T(k)|^2 = \cos^2(\delta_{\rm e}(k)-\delta_{\rm o}(k))$.  The transmission probability is shown in Fig.\ \ref{fig:CI_scat}(b).  As expected, $|T(k)|^2 \rightarrow 0$ in the limit $k \rightarrow 0,$ and $|T(k)|^2 \rightarrow 1$ as $k \rightarrow \infty$.  The smoothness of the Gaussian external potential means that quantum reflections at the trap edges do not play a significant role.  Consequently, $\delta_{\rm e}^{\rm bg}(k) \approx \delta_{\rm o}^{\rm bg}(k)$ and the background transmission profile approaches unity without the characteristic oscillations observed in the case of a square well potential.  In addition to the background transmission profile, we observe twenty-four scattering resonances for which the transmission falls to zero.

The observed scattering resonances are due to interference between the bound two-particle one-hole excitations and the one-particle excitation continuum.  The coupling of the two branches of the excitation spectrum is provided by the inter-atomic pairing, i.e., the matrix elements $V_{qtsr}$ in Eqs.\ (\ref{eqn1_c}) and (\ref{eqn2_c}).  For each resonance, the coupling $V_{qtsr}$ is only non-zero for either the even or the odd continuum wavefunctions, depending on the parity of the $(r,s,t)$ bound wavefunction  [see Eq.\ (\ref{matrix_element})].  Therefore, a particular $(r,s,t)$ configuration modifies either the even or the  odd continuum states, but not both.

The effect on the continuum of an embedded discrete state has been described by Fano \cite{Fano1961}.  Following that treatment, the resonant phase shift $\delta_{\rm e,o}^{\rm R}(k)=\delta_{\rm e,o}(k)-\delta_{\rm e,o}^{\rm bg}(k)$ near each resonance is approximated by
\begin{eqnarray}
\tan \delta_{\rm e,o}^{\rm R}(k) = \frac{\Gamma(k)/2}{\bar{E}_{\nu}-E_{rs}^t -\chi(k)},
\label{res_delta}
\end{eqnarray}
where the resonance width is $\Gamma(k)=2ML|V_{ktsr}|^2/\hbar^2 k$ and the resonance energy is determined by
\begin{equation}
E_{rs}^{t} = \frac{(\hbar k_{rs}^{t})^2}{2M}= E_r+E_s-E_t+V_{rssr}-V_{rttr},
\label{Erst}
\end{equation}
and $\chi(k)=\sum_{q} |V_{qtsr}|^2/(\bar{E}_{\nu}-E_{q})$. Taking $V_{qtsr}$ to be approximately independent of $q$, we find that $\chi(k)=0$.  The resonance energies are well approximated by $\bar{E}_{\nu} = E_{rs}^{t}$, as shown by the vertical lines in Fig.\ \ref{fig:CI_scat}(b).

The energy $E_{rs}^t$ [see Eq.\ (\ref{Erst})] can be interpreted as the energy required to create the bound two-particle one-hole excitation $(r,s,t)$, i.e., to create a spin-up particle in level $r$ and excite a spin-down particle from level $t$ to $s$.  The last two terms in Eq.\ (\ref{Erst}) account for the accompanying change in the pair-wise interaction energy.  When the energy of the incoming projectile particle matches $E_{rs}^t$, the bound excitation $(r,s,t)$ is an allowed intermediate state for the scattering process [see Figs.\ \ref{fig:HF_ground}(b) and \ref{fig:HF_ground}(c) for $(r,s,t)=(7,8,3)$].  The intermediate state is made accessible by the two-body coupling term $V_{qtsr}$ in Eqs. (\ref{eqn1_c}) and (\ref{eqn2_c}).

\section{Mean-field approach \label{sec:BCS}}

The BCS mean-field treatment of pairing and superfluidity in fermionic systems has been extremely successful.  It accurately describes a wide range of systems in  which many-body pair correlations are important.  The tractability of the method is a significant advantage.  However, it is often difficult to conclusively evaluate the validity of the theory.  The scattering problem considered in Sec.\ \ref{sec:HF} is particularly useful for a discussion of the validity of BCS theory because (i) the scattering resonances are extremely sensitive to the ground state properties of the trapped gas, and (ii) a physical interpretation of the resonances is possible allowing the mean-field approach to be investigated in detail.

In the BCS mean-field treatment, we approximate Hamiltonian (\ref{Hamiltonian}) by the Hartree-Fock-Bogoliubov Hamiltonian
\begin{eqnarray}
\hat{H}_{\rm HFB} & =  & \sum_{\alpha} \int \hat{\psi}_{\alpha}^{\dagger} (x) \left[ H_{\rm SP} (x)+U(x)-\mu \right] \hat{\psi}_{\alpha} (x) dx  \nonumber \\
& &+ \int \left[ \Delta(x)  \hat{\psi}_{\uparrow}^{\dagger}(x) \hat{\psi}_{\downarrow}^{\dagger} (x)+H.c. \right] dx ,
\label{Hamiltonian_HFB}
\end{eqnarray}
where the Hartree potential is $U(x)=g \langle\hat{\psi}_{\alpha }^{\dagger}(x)\hat{\psi}_{\alpha}(x) \rangle$ and the pair potential is $\Delta(x)= -g\langle\hat{\psi}_{\uparrow}(x)\hat{\psi}_{\downarrow}(x) \rangle$ \cite{note_HFB}.  The two spin states are equally populated so $U(x)$ is spin independent.  We choose the chemical potential $\mu<0$ so that all of the particles are bound.  The Bogoliubov transformation,
\begin{equation}
\begin{split}
\hat{\psi}_{\uparrow}(x)= & \sum_{n} \left[
u_{n}(x)\hat{\gamma}_{n \uparrow} - v_{n}^{*}(x)\hat{\gamma}^{\dagger
}_{n \downarrow} \right]  \\
\hat{\psi}_{\downarrow}(x)= & \sum_{n} \left[
u_{n}(x)\hat{\gamma}_{n \downarrow} + v_{n}^{*}(x)\hat{\gamma}^{\dagger
}_{n \uparrow} \right] ,
\end{split}
\label{Bog_trans}
\end{equation}
diagonalizes Hamiltonian (\ref{Hamiltonian_HFB}), where the quasiparticle amplitudes $u_{n}(x)$ and $v_{n}(x)$ solve the Bogoliubov-de Gennes equations \cite{deGennes}
\begin{align}
\left[
\begin{array}[c]{cc} {\cal L}(x) & \Delta(x)\\ \Delta^{*}(x) & -{\cal L} (x) \end{array}
\right]  \left[
\begin{array}[c]{c} u_{n}(x)\\ v_{n}(x) \end{array}
\right] = \epsilon_n  \left[
\begin{array}[c]{c} u_{n} (x)\\ v_{n}(x) \end{array}
\right]. \label{bdgeqns}
\end{align}
In Eq.\ (\ref{bdgeqns}), ${\cal L}(x) = H_{\rm SP}(x) +U(x)-\mu$ and $\epsilon_n$ are the quasiparticle energies (taking $\epsilon_n>0$).  The quasiparticle operators $\hat{\gamma}_{n\alpha}^{\dagger}$ and $\hat{\gamma}_{n\alpha}$ obey fermionic commutation relations and the quasiparticle modes are populated according to the Fermi distribution function, i.e., at zero temperature, $\langle \hat{\gamma}_{n\alpha}\hat{\gamma}_{n'\beta}^{\dagger}\rangle=\delta_{nn'}\delta_{\alpha\beta}$ \cite{deGennes}.

\begin{figure}[t]
\includegraphics[width=8.5cm]{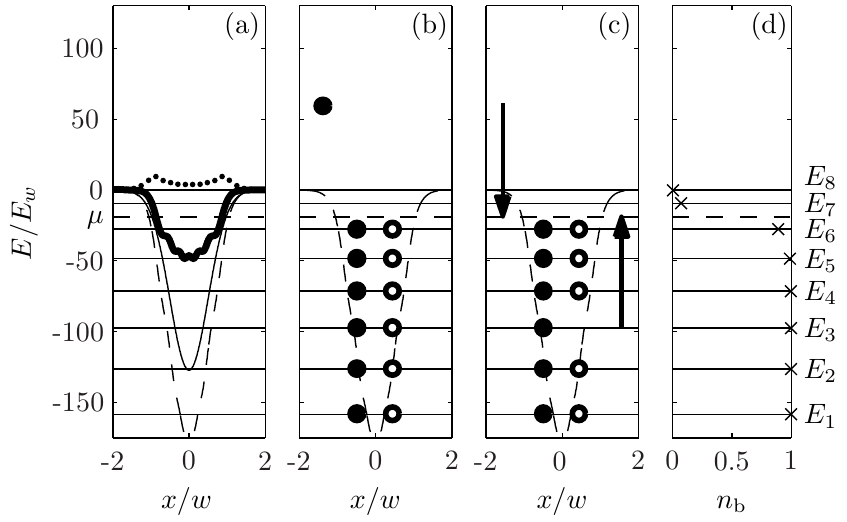}
\caption{(a) The BCS ground state of a trapped Fermi gas.  The curves correspond to (thick) the Hartree potential $U(x)$, (dotted) the pair potential $\Delta(x)$, (thin) the Gaussian external potential $V_{\rm ext}(x)$, and (dashed) the combined potential $\bar{V}(x)=V_{\rm ext}(x)+U(x)$.  (b) A scattering particle excitation, and (c) a bound hole excitation, where the energy level occupation of the ($\bullet$) spin up and ($\circ$) spin down particles is indicated schematically.  The arrows indicate the particle rearrangement between the configurations in panels (b) and (c).  (d)  The ground state occupation of the bound states in the combined potential $\bar{V}(x)$ [see Eqs.\ (\ref{bound}) and (\ref{n_bound})] is indicated by the horizontal component of the markers $\times$.  In all panels, the horizontal lines indicate (dashed) the chemical potential $\mu$ and (solid) the bound state energies $E_n=E_{\rm b}<0$ in the combined potential $\bar{V}(x)$. Parameters are $g=-9.55w E_w$, $V_0=127.32E_w$, and $\mu=-19.10E_w$.}
\label{fig:BCS_ground}
\end{figure}
Figure \ref{fig:BCS_ground}(a) shows the self-consistent BCS mean fields for a trapped Fermi gas.  We have used the same parameters as in Figs.\ \ref{fig:HF_ground} and \ref{fig:CI_scat}.  However, in the mean-field treatment, the ground state is not a particle number eigenstate \cite{deGennes}.  For the parameters used in Fig.\ \ref{fig:BCS_ground}, the average number of particles in each spin state is $\langle \hat{N}_{\alpha}\rangle=5.97$ and the number variance is $\langle \hat{N}_{\alpha}^2 \rangle-\langle \hat{N}_{\alpha} \rangle^2=0.20$.  

To understand the effect of the off-diagonal coupling in the Bogoliubov-de Gennes equations, we first consider Eq.\ (\ref{bdgeqns}) with $\Delta(x) =0$.  We retain the Hartree potential $U(x)$ from the finite $\Delta(x)$ self-consistent solution.  In this case, Eq.\ (\ref{bdgeqns}) reduces to the Schr${\rm \ddot o}$dinger equation
\begin{eqnarray}
\left[ -\frac{\hbar^2}{2M} \frac{d^2}{dx^2}+{\bar V}(x) \right] \psi_n(x)= E_n \psi_n(x),
\label{bound}
\end{eqnarray}
where $\bar{V}(x)= V_{\rm ext}(x)+U(x)$ is the effective confining potential for the many-body system \cite{note_En}.  Equation (\ref{bound}) has discrete bound states $\psi_{\rm b}(x)$ with energy $E_b<0$, and an excitation continuum of both even and odd scattering states $\psi_k^{\rm e,o}(x)$ with energy $E_k=\hbar^2 k^2/2M>0$.  In the particle-hole picture, the excitation spectrum has a particle branch $[u_n^0(x)=\psi_n(x)]$ with energy $\epsilon_n^0=E_n-\mu$ (for $E_n>\mu$) and a hole branch $[v_n^0(x)=-\psi_n(x)]$ with energy $\epsilon_n^0=-E_n+\mu$ (for $E_n<\mu$).  The superscript zero indicates that we are solving Eq.\ (\ref{bdgeqns}) with $\Delta(x)=0$. 

Taking $\Delta(x)$ to be finite introduces coupling between the particle and hole branches of the excitation spectrum.  The quasiparticle excitations then become simultaneously particle- and hole-like to reflect the fact that the pairing interactions can excite particles to energy levels lying above the chemical potential.  Figure \ref{fig:BCS_ground}(d) shows the BCS ground state average particle occupations of the uncoupled bound states $\psi_{\rm b}(x)$, i.e.,
\begin{equation}
n_{\rm b} = \sum_n \left| \int \psi_{\rm b}(x) v_n (x) dx \right| ^2.
\label{n_bound}
\end{equation}
The lowest bound levels are fully occupied but there is a redistribution of particles near the chemical potential, compared to the Hartree-Fock ground state [see Fig.\ \ref{fig:HF_ground}(d)].

To investigate the scattering properties of the system, we compute the even and the odd solutions of the Bogoliubov-de Gennes equations (\ref{bdgeqns}) subject to Neumann boundary conditions.  We then match the asymptotic behavior of the quasiparticle amplitudes to their analytic forms.  The particle-like amplitudes lying in the continuum have the form
\begin{eqnarray}
 \lim_{x \to \pm \infty} u_k^{\rm{e}}(x)& \propto & \cos(kx \mp \delta_{\rm e} (k)) \nonumber \\
 \lim_{x\to \pm \infty} u_k^{\rm{o}}(x)& \propto & \sin(kx \mp \delta_{\rm o} (k)),
\label{u_asympt}
\end{eqnarray}
where $\epsilon_k = \hbar^2 k^2/2M-\mu.$  In the asymptotic limit, the corresponding hole-like amplitudes $v_k^{\rm e,o}(x)$ tend exponentially to zero.  Again we use the R-matrix method \cite{Wigner47} to determine the scattering phase shifts $\delta_{\rm e}(k)$ and $\delta_{\rm o}(k)$ for any $k$.

\begin{figure}[t]
\begin{center}
\includegraphics[width=8.5cm]{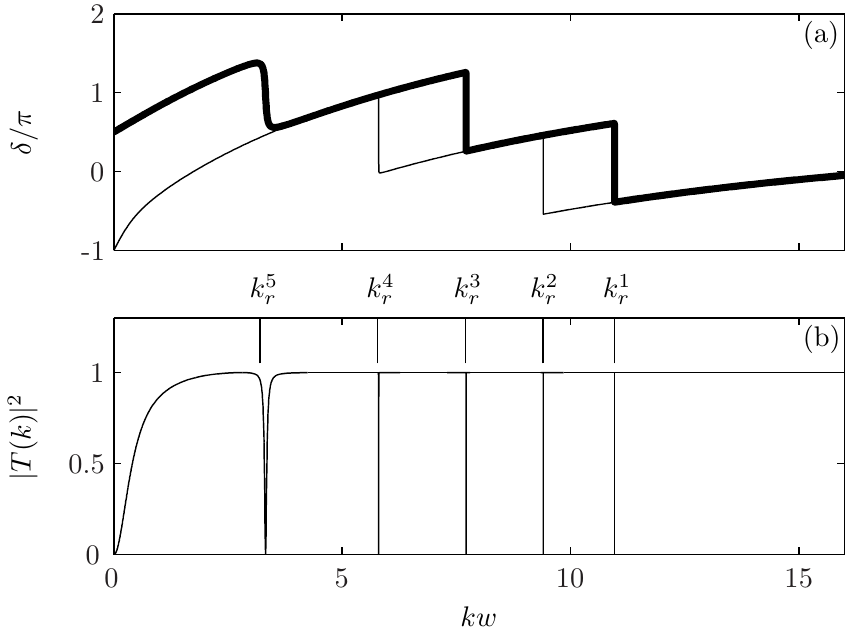}
\caption{(a) Scattering phase shifts and (b) transmission probability for a trapped Fermi gas, calculated using the mean-field method.  The curves in (a) correspond to (thick) the even phase shift $\delta_{\rm e}(k)$ and (thin) the odd phase shift $\delta_{\rm o}(k)$.  The vertical lines in (b) indicate the uncoupled resonance wavevectors $k_r^b$ [see Eq.\ (\ref{res_energies})].  Parameters are $g = -9.55 w E_w$, $V_0 = 127.32 E_w$, and $\mu = -19.10 E_w$.}
\label{fig:BCS_scat}
\end{center}
\end{figure}

The even and the odd phase shifts for the parameters used in Fig.\ \ref{fig:BCS_ground} are shown in Fig.\ \ref{fig:BCS_scat}(a).  The phase shifts are independent of the spin of the projectile particle because the two spin states are equally populated.  We observe a background variation in the phase shifts due to the effective potential $\bar{V}(x)$ [see Eq.\ (\ref{bound})].  We also observe five resonance features that occur alternately in the even and the odd phase shifts and become narrower higher in the continuum.  The transmission probability is shown in Fig.\ \ref{fig:BCS_scat}(b).

The observed scattering resonances are possible because, in the uncoupled system, a scattering particle excitation [e.g., see Fig.\ \ref{fig:BCS_ground}(b)] can have the same quasiparticle energy as a bound hole excitation  [e.g., see Fig.\ \ref{fig:BCS_ground}(c) for ${\rm b}=3$].  In particular, a particle scattering state with quasiparticle energy $\epsilon_k^0=E_k-\mu$ is degenerate with a bound hole excitation with quasiparticle energy $\epsilon_{\rm b}^0=-E_{\rm b}+\mu$ when $\epsilon_k^0=\epsilon_{\rm b}^0$.  In the ordinary particle picture this can be rewritten as  $E_k=E_{\rm r}^{\rm b},$ where
\begin{equation}
E_{\rm r}^{\rm b}=\frac{(\hbar k_{\rm r}^{\rm b})^2}{2M}=-E_b+2\mu.
\label{res_energies}
\end{equation}
If $E_{\rm r}^{\rm b}>0$, the uncoupled bound state is embedded in the particle excitation continuum and, when $\Delta(x)$ is finite, this gives rise to a Fano-type scattering resonance.  Note that the ${\rm b}=6$ bound state in Fig.\ \ref{fig:BCS_ground} does not give rise to a resonance because it does not lie sufficiently low in the trap, i.e., $E_{6}>2\mu.$ 

Using standard techniques \cite{Friedrich}, we find that near a resonance the resonant phase shift is well described by the Fano profile
\begin{eqnarray}
\tan \delta_{\rm e,o}^{\rm R}(k) = \frac{\Gamma_{\rm e,o}(k)/2}{\epsilon_k^0-\epsilon_{\rm b}^0 -\chi_{\rm e,o}(k)}.
\label{res_delta}
\end{eqnarray}
The resonance width is
\begin{equation}
\Gamma_{\rm e,o}(k) = - \frac{2M}{\hbar^2 W_k} Q_{\rm e,o}^2(k) \cos(\delta_e^{\rm bg}(k)-\delta_o^{\rm bg}(k)),
\end{equation}
where $Q_{\rm e,o} (k) = \int \psi_{\rm b}(x) \Delta(x) \psi_k^{\rm e,o}(x) dx$ \cite{note_estates} and the Wronskian $W_k= \psi_k^{\rm o}(x)d\psi_k^{\rm e}(x)/dx - \psi_k^{\rm e}(x)d\psi_k^{\rm o}(x)/dx$ is constant because there is no first order derivative in Eq.~(\ref{bound}) \cite{Arfken}.   For each resonance, the coupling matrix element $Q_{\rm e,o}(k)$ is only non-zero for either the even or the odd continuum wavefunctions, depending on the parity of the bound hole excitation. The smoothness of the Gaussian external potential means that $\delta_{\rm e}^{\rm bg}(k)\approx \delta_{\rm o}^{\rm bg}(k)$ and, therefore, $\cos(\delta_e^{\rm bg}(k)-\delta_o^{\rm bg}(k))\approx 1$.

The resonance energy is determined by solving $\epsilon_k^0=\epsilon_b^0+\chi_{\rm e,o}(k)$, where
\begin{equation}
\chi_{\rm e,o} (k)  = \Theta(k) \pm \frac{M}{\hbar^2 W_k} Q_{\rm e,o}^2(k) \sin (\delta_e^{\rm bg}(k)-\delta_o^{\rm bg}(k)),
\label{chi_eqn}
\end{equation}
and $\Theta(k) =\int  \psi_{\rm b}(x) \Delta(x) G_k(x,s) \Delta(s) \psi_{\rm b}(s) dx ds $.  The upper (lower) sign in Eq.\ (\ref{chi_eqn}) applies if the uncoupled bound hole excitation is even (odd), but $\sin(\delta_e^{\rm bg}(k)-\delta_o^{\rm bg}(k))\approx 0$ and the first term in Eq.\ (\ref{chi_eqn}) dominates.  The Green's function of Eq.~(\ref{bound}) is
\begin{equation}
G_k(x,s) = \frac{M}{\hbar^2 W_k} \left\{
\begin{array}{cc}
\psi_k^{\rm e}(x) \psi_k^{\rm o}(s) - \psi_k^{\rm o}(x) \psi_k^{\rm e}(s), &  x>s \\
\psi_k^{\rm o}(x) \psi_k^{\rm e}(s) - \psi_k^{\rm e}(x) \psi_k^{\rm o}(s), & x<s
\end{array}   \right..
\end{equation}
In the limit $\Delta(x)\rightarrow 0$, $\chi(k)=0$ and the resonances occur for $\epsilon_k^0=\epsilon_{\rm b}^0$, i.e., the quasiparticle energy of the bound hole excitation matches the quasiparticle energy of the scattering particle excitation.  The resonance energy is well approximated by $E_k=E_{\rm r}^{\rm b}$ [see Eq.\ (\ref{res_energies})], as indicated by the vertical lines in Fig.\ \ref{fig:BCS_scat}(b).

The energy $E_{\rm r}^{\rm b}$ can be interpreted as the energy required to excite a bound particle to the chemical potential and create a second particle at the chemical potential with opposite spin.  A scattering resonance occurs near this energy if there is coupling between the scattering state and the intermediate state, where the projectile particle and the bound particle of opposite spin form a pair at the chemical potential.  In general, this intermediate state is forbidden because the chemical potential is not an energy eigenvalue of Eq.\ (\ref{bound}).  However, in the mean-field theory, the pair potential $\Delta(x)$ facilitates pair creation and destruction as if there is a source/sink of atom pairs at the chemical potential, i.e., there is an effective pair condensate at $\mu$.  This allows the projectile particle and a bound particle of opposite spin to be simultaneously removed from the system and the intermediate state for the scattering process is a bound hole excitation with two fewer particles than the scattering state [see Figs.\ \ref{fig:BCS_ground}(b) and \ref{fig:BCS_ground}(c)].  In a number conserving treatment, coupling to this intermediate state would be forbidden because the Hamiltonian (\ref{Hamiltonian}) does not couple states with different numbers of particles.  

We conclude that, in this case, the scattering resonances predicted by the mean-field approach are spurious.  In particular, the mean-field theory does not conserve particle number and this allows for coupling between sectors of the Hilbert space corresponding to different numbers of particles.  We discuss the validity of the BCS mean-field approach in more detail in the following section.

\section{Validity of mean-field theory \label{sec:disc}}

The conventional application of BCS mean-field theory provides a steady state ansatz for the quantum system.  This can be used to compute the expectation values of a variety of physical observables that conserve the total number of particles.  In their seminal paper, Bardeen, Cooper, and Schrieffer state explicitly that they only `$\dots$ for the moment relax the requirement that the wave function describes a system with a fixed number of particles $\dots$' \cite{BCS}.  They go on to explain how the expectation value, of the number conserving Hamiltonian, will yield approximately the same result when evaluated either with the BCS state or with the projection of that state on any eigenstate of the total number of particles, so long as it has a particle number eigenvalue close to the (large) mean particle number in the BCS state.  

Our use of BCS theory is less well justified because, rather than investigating the full number conserving Hamiltonian (\ref{Hamiltonian}) within the BCS ansatz, we address the dynamics governed by the approximate number non-conserving Hamiltonian (\ref{Hamiltonian_HFB}).  Higher order methods based on the number conserving Hamiltonian have been applied to scattering from nuclei (e.g., \cite{Orrigo2006}).  However, in the case of our one-dimensional scattering problem, the spurious scattering resonances described in Sec.\ \ref{sec:BCS} are retained even in such a treatment.

It is well known that the BCS mean-field approach is particularly prone to giving inaccurate results when applied to small systems.  So, it is perhaps not surprising that the BCS treatment predicted spurious resonances in Sec.\ \ref{sec:BCS}.  However, by illustrating in detail the nature of this breakdown of the theory, we can now present arguments for why the BCS method can be expected to give accurate predictions in higher dimensions and for larger systems. 

One of the difficulties with using the BCS mean-field approach to describe the one-dimensional scattering problem is that the chemical potential is not, in general, an allowed energy level in the system.  However, this problem only arises in systems with discrete energy levels.  For example, in two or three dimensions where the energy levels can be highly degenerate, or in a homogeneous system where the energy spectrum is continuous, the chemical potential is far more likely to lie at, or very near to, an available energy level.  Furthermore, the high degeneracy provides a compelling argument for the validity of the mean-field description in the macroscopic limit, as discussed below.

\begin{figure}[t]
\begin{center}
\includegraphics[width=4.5cm]{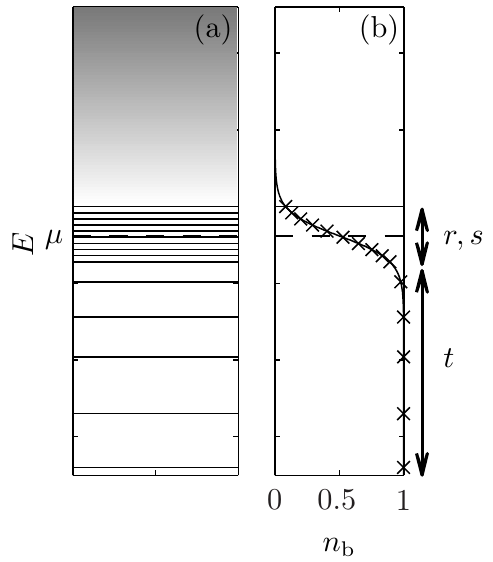}
\caption{(a) The energy level diagram for a generic quantum system for which the energy levels near the chemical potential are highly degenerate.  (b) The average energy level population $n_{\rm b}$ that could be expected from a  BCS mean-field calculation.  In both panels, the dashed horizontal line indicates the chemical potential $\mu$.}
\label{fig:illustration}
\end{center}
\end{figure}

Figure \ref{fig:illustration}(a) represents the excitation spectrum for a generic quantum system where the energy levels are highly degenerate near the chemical potential.  Applying the BCS mean-field theory to such a system would give an energy level average population $n_{\rm b}$ that varies slowly across the quasi-continuum, as indicated in Fig.\ \ref{fig:illustration}(b).  Calculating the scattering properties of the system, in the mean-field approach, we would expect to find scattering resonances at energies $E_k\approx E_{\rm r}^{\rm b}$ for each sufficiently low lying energy level (i.e., for $E_{\rm b}<2\mu$).  To determine whether this is reasonable, we could alternatively consider this system in the spirit of the configuration-interaction approach of Sec.\ \ref{sec:HF}.  In this treatment we would expect there to be a large number of resonances due to the many possible $(r,s,t)$ configurations.  In particular, the resonances for a particular value of $t$ would have similar resonance energies and, taking the continuum limit for the $r$ and $s$ levels, the many resonances would overlap and we would predict essentially the same result as in the BCS mean-field treatment, i.e., a single resonance for every low lying bound state $t$.  To verify this quantitatively, it would be necessary to show that the operator $\hat{a}_{r\uparrow}^{\dagger}\hat{a}_{s\downarrow}^{\dagger}$ in Eq.\ (\ref{Q_excite}) could be replaced by a complex number that was approximately independent of $r$ and $s$ in the quasi-continuum.  Therefore, Eq.\ (\ref{Q_excite}) could be re-interpreted in the spirit of the Bogoliubov transformation (\ref{Bog_trans}).

To rephrase the argument, consider a group of ${\cal N}$ states near the chemical potential that are on average half filled.  In this case the states near the chemical potential are mutually coupled by the many-body pairing and the corresponding eigenstates can be assumed to be essentially symmetric in terms of the population of each single particle state.  The probability of transferring the projectile particle, and a low-lying bound particle of opposite spin, to any of the half-filled levels is then amplified by a combinatorial factor increasing with $\cal N$, similar to the collective spontaneous emission (Dicke superradiance) of light from a symmetrically excited atomic medium \cite{Dicke}.   The collective enhancement would be maximum at exactly half filling and the intermediate states populating $r$ and $s$ levels near $\mu$ would yield one strong resonance for every level $t$, in agreement with the BCS mean-field prediction.  Additional resonances due to unpopulated $r$ and $s$ levels lying well above the chemical potential [as in Fig.\ \ref{fig:CI_scat}] do not benefit from the collective enhacement associated with the half filling of levels and would be comparatively unimportant in the macrosopic limit.  This justifies the use of the self-consistent mean-field treatment as a symmetry breaking approach in the same spirit that the semiclassical approximation with a mean collective dipole is applied to the description of optical superradiance. 

\section{Conclusion \label{sec:conclusion}}

We have considered single-particle scattering from a trapped two-component degenerate Fermi gas when the projectile particle is identical to one of the confined species. Our theoretical treatment is based on a configuration-interaction approach and we predict Fano-type scattering resonances that are possible because of inter-atomic pairing.  The scattering resonances are sensitive to the ground state properties of the trapped Fermi gas and we have described the key features of the scattering resonances quantitatively.

We have also presented a BCS mean-field approach to the scattering problem and have shown that the non-number conservation of the theory leads to spurious scattering resonances.  We have described in detail the breakdown of BCS theory for this case but have argued that in macroscopic systems where the energy spectrum is highly degenerate, or continuous, the BCS theory may be expected to give accurate results.  Furthermore, we have suggested that the BCS mean-field approximation can be interpreted as an effective semi-classical representation of superradiance in the system, and this presents an interesting avenue for further research.  

The scattering resonances we predict are relevant for a range of reflection and transmission experiments in quasi-one-dimensional geometries (e.g., \cite{Kinoshita06, Billy08, Chin06}).  Transport studies of different junction interfaces are of particular interest for developing electronic devices \cite{Demers71, Kalenkov07,Das08}.  For example, the energy sensitivity of the scattering resonances could allow for energy filtering and, in the case of spin imbalance, spin filtering may also be possible.  Recent work has also shown that the scattering properties of normal-superfluid interfaces have implications when considering the thermodynamics of Fermi gases \cite{Schaeybroeck07}.

\begin{acknowledgments}
The authors thank Stefan Rombouts for helpful discussions, and N. N. acknowledges financial support from the Danish Natural Science Research Council.
\end{acknowledgments}



\begin{thebibliography}{25}

\bibitem{Inouye98} S.\ Inouye, M.\ R.\ Andrews, J.\ Stenger, H.\ J.\ Miesner, D.\ M.\ Stamper-Kurn, and W.\ Ketterle, Nature \textbf{392}, 151 (1998).

\bibitem{Jochim03} S.\ Jochim, M.\ Bartenstein, A.\ Altmeyer, G.\ Hendl, S.\ Riedl, C.\ Chin, J.\ Hecker Denschlag, and R.\ Grimm, Science \textbf{302}, 2101 (2003).

\bibitem{Regal03} C.\ A.\ Regal, C.\ Ticknor, J.\ L.\ Bohn, and D.\ S.\ Jin, Nature \textbf{424}, 47 (2003).

\bibitem{Zwierlein05} M.\ W.\ Zwierlein, J.\ R.\ Abo-Shaeer, A.\ Schirotzek, C.\ H.\ Schunck and W.\ Ketterle, Nature \textbf{435}, 1047 (2005).

\bibitem{Chin04} C.\ Chin, M.\ Bartenstein, A.\ Altmeyer, S.\ Riedl, S.\ Jochim, J.\ Hecker Denschlag, and R.\ Grimm, Science \textbf{305}, 1128 (2004).

\bibitem{R&S} P.\ Ring and P.\ Schuck, \textit{The Nuclear Many-Body Problem} (Springer-Verlag, Berlin, 2004).

\bibitem{Bruun98} G.\ M.\ Bruun and K.\ Burnett, Phys.\ Rev.\ A \textbf{58}, 2427 (1998).

\bibitem{note_CImodes} We assume that the continuum modes do not contribute to the bound component of the excitation, i.e., that the Hartree-Fock method gives a good description of the single-particle bound state energy levels of the system.

\bibitem{Wigner47} E.\ P.\ Wigner and L.\ Eisenbud, Phys.\ Rev.\ \textbf{72}, 29 (1947).

\bibitem{Poulsen03} U.~V.~Poulsen and K.~M{\o}lmer, Phys.\ Rev.\ A \textbf{67}, 013610 (2003).

\bibitem{Grupp06} M.\ Grupp, G.\ Nandi, R.\ Walser, and W.\ P.\ Schleich, Phys.\ Rev.\ A \textbf{73}, 050701(R) (2006).

\bibitem{Fano1961} U.\ Fano, Phys.\ Rev.\ \textbf{124}, 1866 (1961).

\bibitem{note_HFB} The Hartree-Fock-Bogoliubov ground state is used when evaluating matrix elements in the mean-field approach.  Therefore, the Hartree potential $U(x)$ in the Hartree-Fock-Bogoliubov treatment is not identical to the Hartree potential $W(x)$ in the Hartree-Fock treatment.

\bibitem{deGennes} P.\ G.\ de Gennes, \textit{Superconductivity of metals and
alloys} (W.\ A.\ Benjamin Inc., New York, 1966).

\bibitem{note_En} The energy eigenvalues $E_n$ of Eq.\ (\ref{bound}) are not quantitatively identical to the energy eigenvalues of Eq.\ (\ref{HF_basis}).

\bibitem{Friedrich} H.\ Friedrich, \textit{Theoretical Atomic Physics} (Springer-Verlag, Berlin, Heidelberg, 1991).

\bibitem{note_estates} We have chosen $\Delta(x)$ and the eigenstates $\psi_n(x)$ of Eq.\ (\ref{bound}) to be real.

\bibitem{Arfken} G.\ B.\ Arfken and H.\ J.\ Weber, \textit{Mathematical Methods for Physicists  - 4th ed.} (Academic Press, Inc., California, 1995).

\bibitem{BCS} J.\ Bardeen, L.\ N.\ Cooper, and J.\ R.\ Scrieffer, Phys.\ Rev.\ \textbf{108},
1175 (1957).

\bibitem{Orrigo2006} S.\ E.\ A.\ Orrigo, H.\ Lenske, F.\ Cappuzzello, A.\ Cunsolo, A.\ Foti, A.\ Lazzaro, C.\ Nociforo, and J.\ S.\ Winfield, Phys.\ Lett.\ B \textbf{633}, 469 (2006).

\bibitem{Dicke} R.\ H.\ Dicke, Phys.\ Rev.\ \textbf{93}, 99 (1954).

\bibitem{Chin06} J.\ K.\ Chin, D.\ E.\ Miller, Y.\ Liu, C.\ Stan, W.\ Setiawan, C.\ Sanner, K.\ Xu, and W.\ Ketterle, Nature \textbf{443}, 961 (2006).

\bibitem{Kinoshita06} T.\ Kinoshita,  T.\ Wenger, and D.\ S.\ Weiss, Nature  \textbf{440}, 900 (2006).

\bibitem{Billy08} J.\ Billy, V.\ Josse, Z.\ Zuo, W.\ Gu\'erin, A.\ Aspect, and P.\ Bouyer, Ann.\ Phys.\ Fr.\ \textbf{32} 17 (2007).

\bibitem{Das08} S.\ Das, S.\ Rao, and A.\ Saha, Europhys.\ Lett.\ \textbf{81}, 67001 (2008).

\bibitem{Kalenkov07} M.\ S.\ Kalenkov and A.\ D.\ Zaikin, Phys.\ Rev.\ B \textbf{75} 172503 (2007).

\bibitem{Demers71} J.\ Demers and A.\ Griffin, Canadian J.\ of Phys.\ \textbf{49}, 285 (1971).

\bibitem{Schaeybroeck07} B.\ V.\ Schaeybroeck and A.\ Lazarides, Phys.\ Rev.\ Lett.\ \textbf{98}, 170402 (2007).

\end{thebibliography}
\end{document}